\documentclass[preprint]{elsarticle}

\usepackage{lineno,hyperref}
\modulolinenumbers[5]

\journal{Computers and Fluids}

\usepackage{amssymb}

\usepackage{amsmath}
\usepackage{graphicx}
\usepackage{bm}
\usepackage{MnSymbol}
\usepackage{caption}
\usepackage{subcaption}

\begin{document}

\begin{frontmatter}

\title{Optimal sensor placement using machine learning}

\author{R. Semaan}
\address{Institute of fluid mechanics, Technische Universit\"at Carolo-Wilhelmina zu Braunschweig, Germany.}
\ead{r.semaan@tu-bs.de}


\begin{abstract}
A new method for optimal sensor placement based on variable importance of machine learned models is proposed.
With its simplicity, adaptivity, and low computational cost, the method offers many advantages over existing approaches.
The new method is implemented on the flow over an airfoil equipped with a Coanda actuator. 
The analysis is based on flow field data obtained from 2D unsteady Reynolds averaged Navier-Stokes (URANS) simulations with different actuation conditions. 
The optimal sensor locations is compared against the current de-facto standard of maximum POD modal amplitude location,
 and against a brute force approach that scans all possible sensor combinations.
The results show that both the flow conditions and the type of sensor have an effect on the optimal sensor placement, 
whereas the choice of the response function appears to have limited influence.
\end{abstract}

\begin{keyword}
Machine learning \sep optimal sensor placement \sep flow control
\end{keyword}

\end{frontmatter}

\section{Introduction} \label{sec:Intro}
The number of sensors and their placement are critical for accurate model predictions and for closed-loop control applications. 
Optimal sensor placement reduces the instrumentation cost and increases the efficiency of the state estimators. 
All known methods to determine the optimal sensor placement rely on either proper orthogonal decomposition (POD) \cite{Cohen2004,Willcox2006,Yildirim2009}, 
or on complicated optimization schemes \cite{Xu1994,Padula1999}.
The current study proposes a new general approach to determine the optimal sensor placement using variable importance of machine learned models. 
The new method circumvents the necessity for POD and for optimization, thereby reducing the complexity and the computational cost.
The approach is applied on a circulation control wing under three different forcing conditions using two sensor types (pressure and shear). 

Optimal sensor placement for structural health monitoring has been repeatedly studied (e.g. Xu et al. \cite{Xu1994} , Padula \& Kincaid \cite{Padula1999}, and Papadopulos \& Garcia \cite{Papadopoulos1998}).
Most approaches start by simulating the structure using finite element methods (e.g. Guratzsch et al. \cite{Guratzsch2005}), 
or by modeling it (e.g.  Fijani \& Vatan \cite{Fijany2005}, Lui \& Tasker \cite{Liu1995}).
The optimal sensor placement is then usually determined with the aid of an optimization scheme (e.g. Xu et al. \cite{Xu1994}, Hamilton \cite{Hamilton2007}).
In principle, these methods are transferable to the fluid dynamics field, but due to their complexity (e.g. solving the algebraic Riccati equation) and numerical costs, 
they have only been implemented on simplified flows (e.g. Burns \& Belinda \cite{Burns1994}, Allan et al. \cite{Allan2000}).
Recently, Brunton et al. \cite{Brunton2013} used compressive sensing to infer optimal sensor placement for image recognition.
Despite its innovative approach, this method also relies on POD in its first step.

The proper orthogonal decomposition has been widely used for a broad range of applications, such as image processing, signal analysis, data compression, and more recently optimal control.
Essentially, POD is a linear procedure that seeks to find a deterministic vector field which has the maximum projection on a random vector field in a mean square sense.
The deterministic functions constitute the orthogonal basis, which are determined as the solutions of an integral eigenvalue problem known as a Fredholm equation. 
These eigenfunctions are optimal in terms of representation of the energy present within the data.

A variety of POD-based methods have been used to infer optimal placement of sensors in fluid flows.
Cohen et al. \cite{Cohen2004} used a heuristic approach to determine the sensor locations by placing them at the extrema of the POD modes,
whereas Willcox \cite{Willcox2006} and Yildirim et al. \cite{Yildirim2009} used Gappy POD.
Gappy POD, developed by Everson \& Sirovich \cite{Everson1995}, can reconstruct incomplete datasets by solving an additional linear system.
The Gappy POD results were similar to the heuristic approach, and the optimum sensor locations almost coincided with the modes extrema.
A similar approach was also used by Mokhasi and Rempfer \cite{Mokhasi2004}.
Kumar et al. \cite{Kumar2014} used linear stochastic estimation (LSE) to infer the sensor positions by minimizing the error between the LSE predictions and the reference POD mode amplitudes.
Their approach contained large uncertainties caused by the limited sensitivity of the LSE predictions towards sensor positions.
As previously mentioned, all existing approaches for engineering applications rely on POD for their sensor placement.
However, proper orthogonal decomposition is sometimes difficult (e.g. very large mesh from LES) or impossible (no spatial field data, or steady flow) to compute.
Moreover, using a finite number of POD mode amplitudes as state estimators is sometimes inaccurate, 
as the relevant aerodynamic properties are not always linearly related to the small subset of selected modes. 
This issue is compounded by the fact that all POD-based methods require a minimum number of sensors, which must be always equal or larger than the total number of considered modes.
This means, for a flow with a shallow modal energetic distribution, one is compelled to use a large number of sensors.

Beside the methodological constraints, all the aforementioned studies failed to investigate the effect the sensor type 
(e.g. pressure sensor, or shear stress sensor) has on the placement.
In addition, the optimal sensor distribution was only determined for a single flow condition,
and thus does not capture the likely variations in the optimal sensor positions with varying flow/actuation conditions. 

\begin{figure}
\centering
\begin{subfigure}{.5\textwidth}
  \centering
  \includegraphics[width=.9\linewidth]{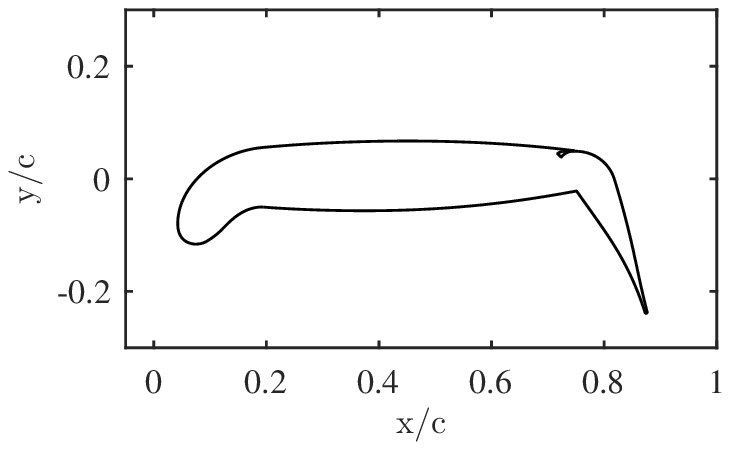}
  \caption{}
  \label{fig:airfoil}
\end{subfigure}%
\begin{subfigure}{.5\textwidth}
  \centering
  \includegraphics[width=.9\linewidth]{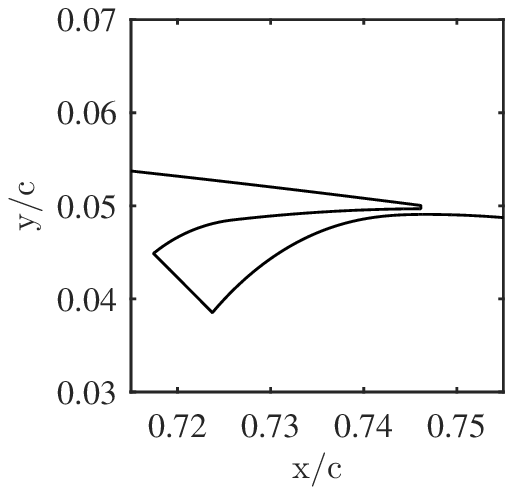}
  \caption{}
  \label{fig:airfoilZoom}
\end{subfigure}
\caption{(a) The modified DLR-F15 high-lift configuration, and (b) a close-up of the actuation duct.}
\label{fig:AirfoilGeo}
\end{figure}

This study introduces a new machine learning based approach to identify optimal sensor positions for a range of conditions and using two different type of sensors.
The method is implemented on three URANS numerical simulations of a circulation control wing  under different forcing conditions. 
The circulation control wing is presented in figure \ref{fig:AirfoilGeo}, where the highly deflected Coanda flap and the actuation duct (close-up in figure \ref{fig:airfoilZoom}) can be seen.
However, efficiency requirements demand that the lift gained through the use of circulation control be as large as possible in comparison to the momentum coefficient of the blown jet, 
which is usually acquired by engine bleed. 
This ratio is referred to as the lift gain factor.
An increase in the lift gain factor is achieved through periodic blowing (e.g. Jones et al. \cite{Jones02}).

This work is part of a larger project that aims to improve the actuation efficiency of airfoils with blown flaps.
Beside the numerical and theoretical investigations, the research project entails a water tunnel experiment.
Periodic excitation of the Coanda jet will be performed using custom-made high-frequency proportional valves,
that allow independent control of the steady and the unsteady blowing components.
This type of actuation enables fine tuning of the jet actuation frequency, amplitude and mean momentum. 
This flexibility coupled with high sensing capability from optimally positioned surface-mounted pressure and shear stress sensors enables closed-loop control, 
which promises higher lift gain factors.
%
%
%
\section{\label{Sec:method}Overview of method}
\begin{figure}
\centering
\includegraphics[width=8.5 cm]{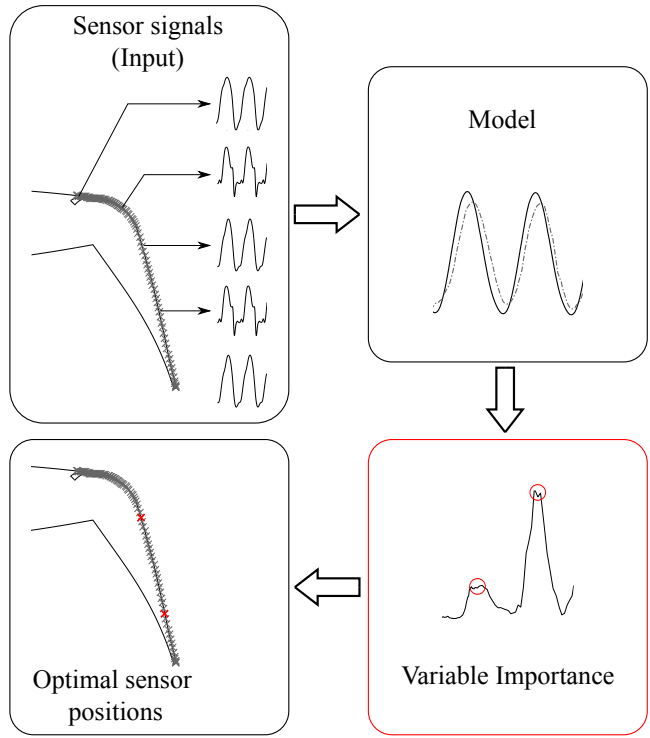}
\caption{\label{fig:algorithm} Schematic summarizing the optimal sensor placement approach using variable importance of machine learning models.}
\end{figure}

Machine learning encompasses many data-driven algorithms and methods.
A supervised learning algorithm takes a set of input data and corresponding outputs (responses), and trains a model to generate predictions for the response to new input data.
For this work, the input data are the instantaneous signals (pressure or wall shear stress) from the airfoil surface,
whereas the response could be any function characterizing the instantaneous flow state.
The optimal sensors are determined as the input signals (from their corresponding locations) with the highest relevance to the trained model.
For this study, the random forests \cite{Breiman2001} algorithm was employed.
It is important to note that the proposed method can be implemented using any supervised machine learning regression algorithm.
Random forests was selected due to its lower out-of-bag (OOB) error and its inherent OOB randomization, 
which is convenient for variable importance (see $\S$ \ref{sec:SensorPlacement}).
The out-of-bag error is the prediction error of machine learning models that utilize bootstrap aggregation (i.e., sampling with replacement).
Specifically, it is the mean prediction error of the data left out from the bagging procedure.
This section describes the overall approach used to predict the flow and to determine the optimal sensor positions.
It is schematically illustrated in figure \ref{fig:algorithm}.
All algorithms and data processing are implemented using Statistics and $\text{Machine Learning Toolbox}^{\text{TM}}$ from Matlab.
\subsection{Random forests}
The random forests algorithm belongs to the family of Classification and Regression Trees (CART), 
where a decision tree with binary splits that maximizes the information gain is constructed.
Details on the algorithm can be found in many publications on the subject (e.g. Breiman \cite{Breiman2001}, Hastie et al. \cite{Hastie2013}).
Only an overview is provided here.

\begin{figure}
\centering
\includegraphics[width=7 cm]{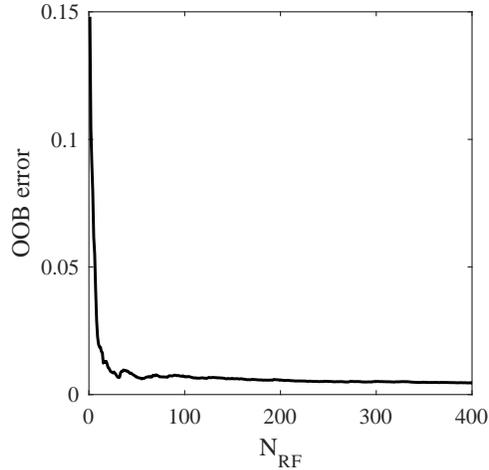}
\caption{\label{fig:OOBError_Trees} Out-of-bag error versus the number of trees, $N_{RF}$.}
\end{figure}

Random forests is an ensemble method that builds multiple regression trees by repeatedly resampling training data with replacement, and averaging the results.
Each tree model predicts the response variable by learning simple if-then-else decision rules from the data.
The random sampling has two main benefits.
First, it increases tree diversity, and thus improves robustness of the prediction.
Second, it is less prone to overfitting, since at each split only a subset of the features is used.
Overfitting is when models reproduce the training data very well but poorly predict unseen data. 
Since there is no limitation on the tree depths in random forests, the algorithm requires only one parameter to set: the number of decision trees, $N_{RF}$.
The necessary number of decision trees can be evaluated by tracking the out-of-bag error,
which is directly obtained from the bootstrapping procedure.
In random forests, there is no need for cross-validation. 
Each tree is constructed using a different bootstrap sample with one-third of the samples left out (referred to as out-of-bag) from each tree.
Figure \ref{fig:OOBError_Trees} shows the regression error rate as function of the number of trees for one of the models in this paper.
As the forest grows larger, the accuracy increases but with diminishing returns.
The error rate decreases drastically between 1 and 30, followed by some fluctuations and small gradual improvement as $N_{RF}$ increases.
As the difference in training times between the various sized trees is negligible, $N_{RF}$ is set to 200 for good accuracy.

\subsection{Data sampling and acquisition}
The first step in supervised machining learning is to generate a suitable training dataset.
Here we have used numerical simulations due to the ease of extracting data everywhere in the field and especially over the airfoil model surface.
Since this study's ultimate objective is to perform closed-loop control,
it is necessary to investigate the flow and to determine the optimal sensor locations under different flow conditions.
Details of the numerical simulations and the different test cases are presented in section \ref{Sec:Num}.

\subsection{Input formulation}
\begin{figure}
\centering
\includegraphics[width=7 cm]{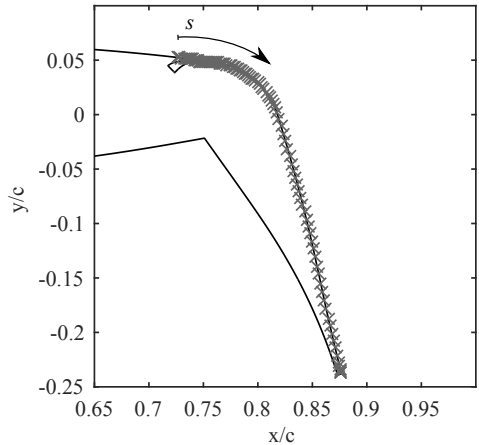}
\caption{\label{fig:AllSensorLoc} All possible sensor locations over the highly deflected flap.
Also shown is the curvilinear coordinate $s$ along the flap surface with its origin at the first possible sensor location.}
\end{figure}
The input training data for the current machine learned models are 96 instantaneous signals.
In order to obtain practical results in an engineering sense, the sensor locations are restricted to the model surface,
i.e. no sensors are allowed in the flow.
The region over the Coanda flap, shown in figure \ref{fig:AllSensorLoc}, is the most sensitive to changes occurring in the wake and to changes originating from the Coanda jet. 
The range of possible sensor locations is therefore further restricted to the flap upper surface and to a small region over the jet exit slot. 
All the 96 considered possible sensor locations are shown by the gray marks in figure \ref{fig:AllSensorLoc}. 
Also shown is the curvilinear coordinate $s$ along the flap surface with its origin at the first possible sensor location.
Two types of input from these sensors is considered: pressure $p$, and skin friction coefficient $C_f$.
The skin friction coefficient is defined as
\[
 C_f=\frac{\tau}{1/2 \rho U^2_\infty}\>,
\]
where $U_\infty$ is the free-stream velocity, and $\tau$ is the wall shear stress.

\subsection{Response function selection}\label{sec:ResponseFunction}
The state of a flow can be characterized through a multitude of metrics, which constitute the response functions for this machine learning problem.
The choice of the metric depends on the problem's objectives and on the data's availability.
For the current flow, several response functions that characterize the state of the flow are conceivable.
In the following, three response functions are presented.
\bibliographystyle{elsarticle-num}

\subsubsection{Proper orthogonal decomposition}
\begin{figure}
\centering
\includegraphics[width=10 cm]{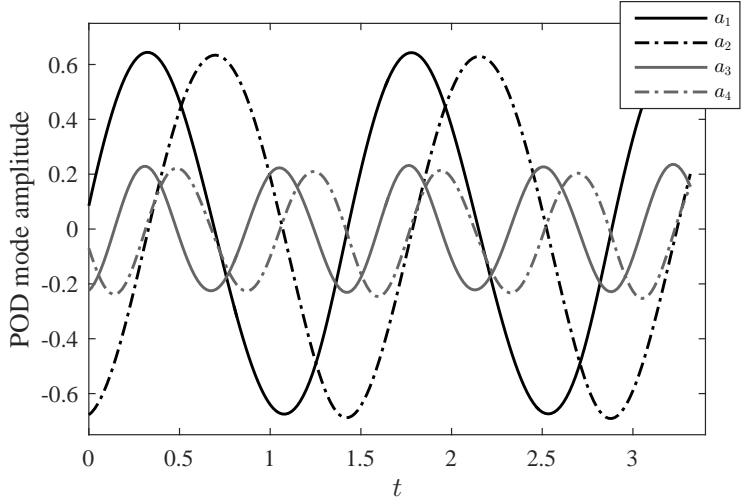}
\caption{\label{fig:ModeAmplitudes}The first four POD mode amplitudes for the unactuated case.}
\end{figure}
\begin{figure}
\centering
\begin{subfigure}{.5\textwidth}
  \centering
  \includegraphics[width=.9\linewidth]{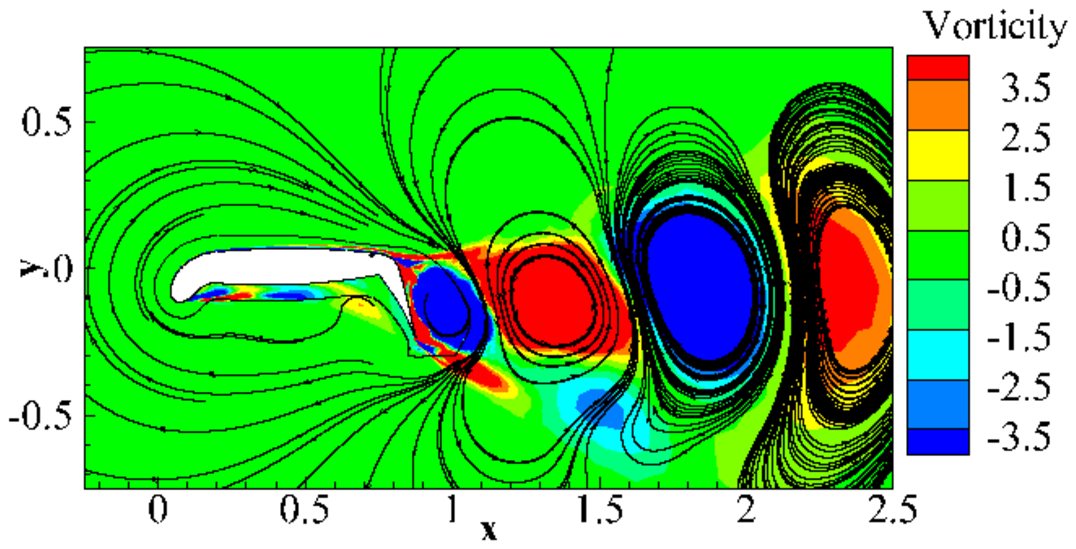}
  \caption{First POD mode}
  \label{fig:Mode1}
\end{subfigure}%
\begin{subfigure}{.5\textwidth}
  \centering
  \includegraphics[width=.9\linewidth]{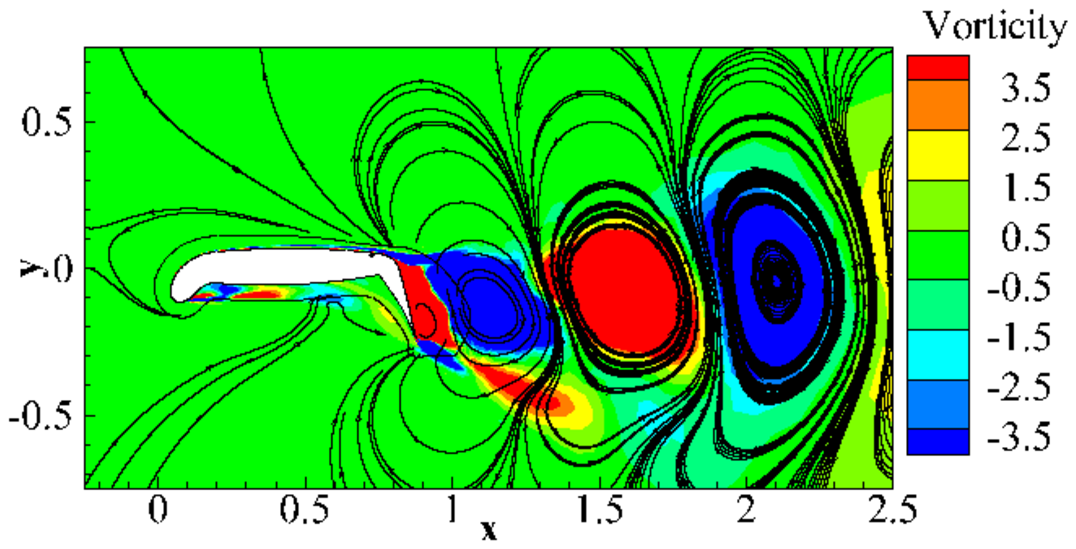}
  \caption{Second POD mode}
  \label{fig:Mode2}
\end{subfigure}
\\
\centering
\begin{subfigure}{.5\textwidth}
  \centering
  \includegraphics[width=.9\linewidth]{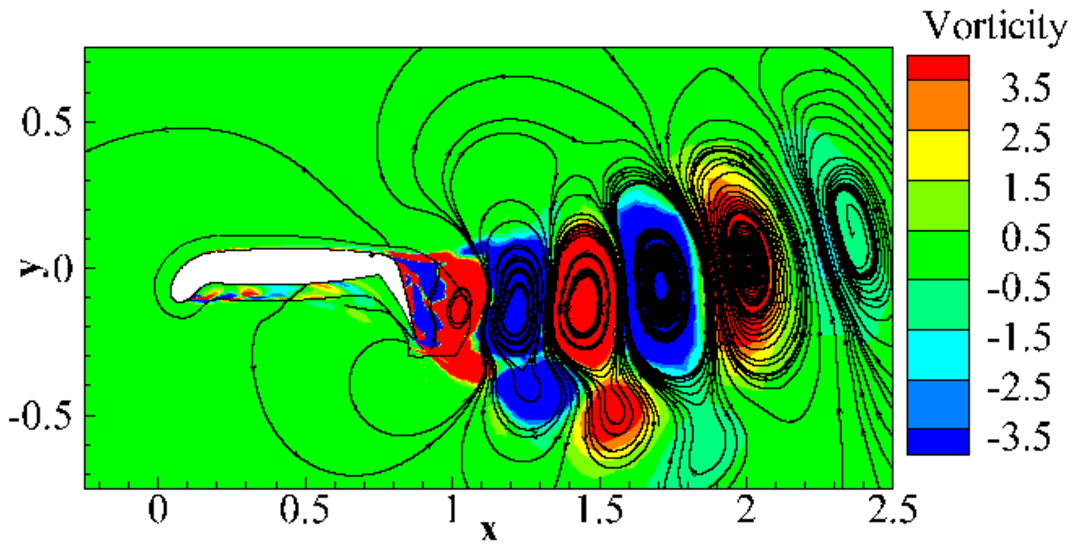}
  \caption{Third POD mode}
  \label{fig:Mode3}
\end{subfigure}%
\begin{subfigure}{.5\textwidth}
  \centering
  \includegraphics[width=.9\linewidth]{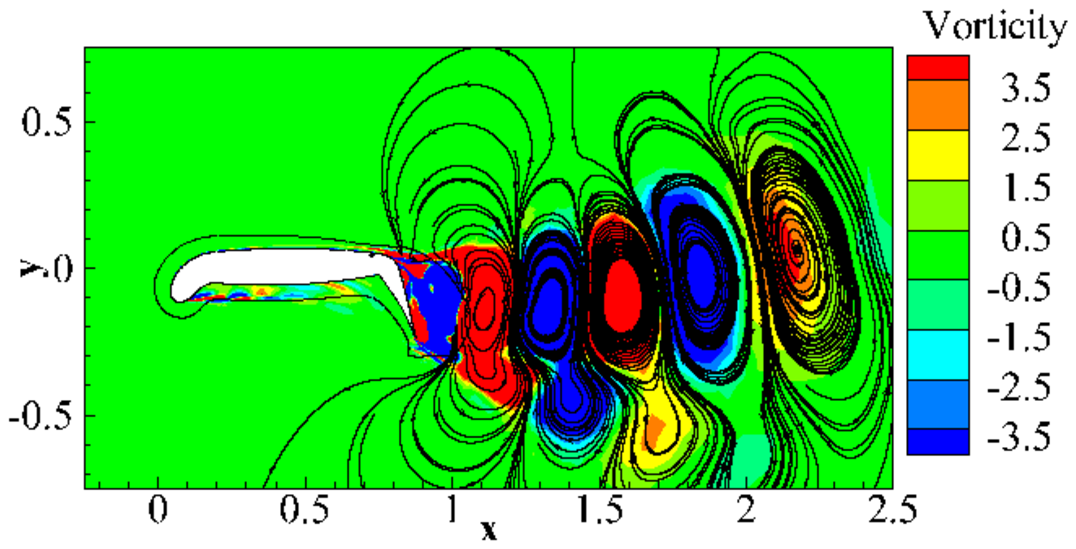}
  \caption{Fourth POD mode}
  \label{fig:Mode4}
\end{subfigure}
\caption{The first four POD modes visualized by the vorticity fields and vortex lines.}
\label{fig:Modes}
\end{figure}

Proper orthogonal decomposition (POD) methods are powerful tools for data analysis aimed at obtaining low-dimensional approximate descriptions of a high-dimensional problem. 
POD is the most optimal modal decomposition in the sense that no other decomposition of the same order captures an equivalent amount of kinetic energy. 
POD of a 2D flow assumes that the velocity fluctuations $\tilde{u}$ can be decomposed as
\begin{equation}
    \label{eq:Linear_Decomp}
    \tilde{u}(x,y,t) = \sum_{k=1}^N a_k(t)\>{\phi}_k(x,y)\,,
\end{equation} 
where $\phi_k$ are the POD modes, $a_k$ are their corresponding amplitudes, and $N$ is the total number of modes. 
The method of snapshots as introduced by Sirovich \cite{sirovich:03a} is used to perform the decomposition. 
Here, the mode amplitudes are first determined from the solution of the eingenvalue problem
\begin{equation}
    \label{eq:EigenProblem}
    C\>A = \lambda A\,,
\end{equation} 
where $\lambda$ is the eigenvalues diagonal matrix, and $A$ is the eigenvector matrix containing the mode amplitudes. 
The cross-correlation matrix $C$ is defined as
\begin{equation}
 \label{eq:CorrMatrix}
 C_{ij} = \frac{1}{N} \langle \tilde{u}(x,y,t_i) \cdot \tilde{u}(x,y,t_j) \rangle\>,
\end{equation}
with $\langle\> \cdot \> \rangle$ the spatial averaging operator.
The POD modes are then determined by projecting the mode amplitudes onto the snapshots and normalizing
\begin{equation}
 \label{eq:Modes}
 \phi_k(x,y) = \frac{1}{N\lambda_k} \sum_{i=1}^{N} a_k^i\>u(x,y,t_k)\>. 
\end{equation}

The POD mode amplitudes represent the temporal evolution of the modes and by extension of the flow.
The POD mode amplitudes, or linear combinations thereof ($\sum_{i=1}^N a_k^i$), are therefore chosen as response functions.
In this study, only the first four modes are considered, which capture more than 90\% of the total kinetic energy for all three test cases.
Their mode amplitudes are shown in figure \ref{fig:ModeAmplitudes} for the unactuated case,
where mode pairing (mode 1--2, and mode 3--4) typical for highly shedding flows can be observed.
The corresponding first four modes for this unactuated case are presented in figure \ref{fig:Modes}, visualized by their vorticity fields and vortex lines.
The first two POD modes represent von K\'{a}rm\'{a}n vortex shedding. 
The third and fourth modes resolve the first harmonic of the dominant shedding frequency. 
Proper orthogonal decomposition was performed using the xAMC software package \cite{xAMC4}.



\subsubsection{Lift coefficient}
The lift coefficient is usually the most relevant parameter in aircraft aerodynamics.
It is therefore the ideal metric for a response function.
The lift coefficient is defined as
\[
 C_l=\frac{L}{1/2 \rho\> U^2_\infty\>c}\>,
\]
where $L$ is the lift force, and $c$ the airfoil chord length with the retracted flap.
The lift coefficient can be experimentally acquired from a balance or from the surface pressure distribution.
Numerically, it is directly integrated from the solution.

\subsubsection{Streamline tracking}
\begin{figure}
\centering
\includegraphics[width=7 cm]{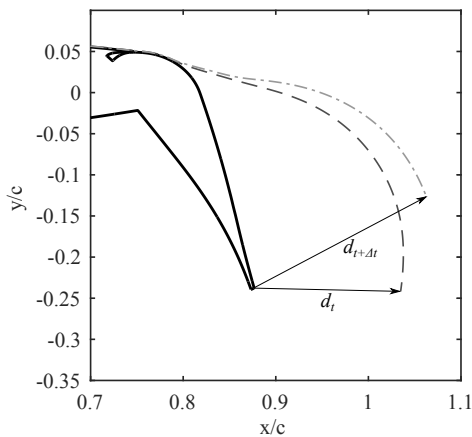}
\caption{\label{fig:Streamlines}Schematic illustrating streamline tracking.}
\end{figure}

The third tested metric that reflects the state of the flow is the streamline distance $d$.
The process is illustrated in figure \ref{fig:Streamlines}.
Similar to experimentally tracking the tip of a long weightless tuft, 
the tip of a constant-length streamline is tracked in real time by computing the distance $d$ between the airfoil trailing edge and the streamline tip.
The streamline origin and length are chosen such that the distance $d$ is zero for an attached state.
The streamline origin is therefore located very near the airfoil surface upstream of the blowing slit.
Changes in $d$ reflect the state of the wake, where, for example, large values and high fluctuations indicate a separated flow.

\subsection{Optimal sensor placement}\label{sec:SensorPlacement}
The optimal sensor positions are simply determined as the most important input variables of the machine learned model.
Since each sensor signal is treated as an input variable, the most important variables are consequently the optimal sensors.
In data mining applications the input variables are rarely equally relevant.
Often only few of them affect the response; and the rest can be excluded.
Several methods can be used to rank the variable importance, such as the Gini splitting index and the OOB randomization.
In this study, the OOB randomization method was selected for its ease of implementation as a byproduct of random forests,
and for its well documented accuracy for variables of the same type \cite{Strobl2007}.
The OOB randomization yields a measure of importance for each predictor variable. 
This measure is computed as the increase in prediction error for any variable as their values are permuted across the out-of-bag observations.
This process is repeated for every tree, then averaged over the entire ensemble and divided by the standard deviation over the entire ensemble.

\begin{figure}
\centering
\includegraphics[width=7 cm]{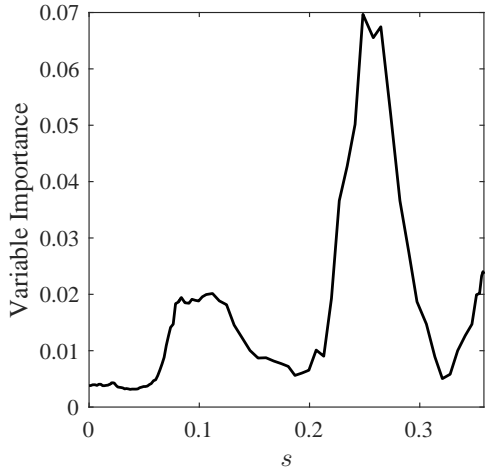}
\caption{\label{fig:VariableImportance} Variable (sensor) importance distribution over the flap to predict $a_1$ for the unactuated case.}
\end{figure}

An example of the variable (sensor) importance distribution over the flap is presented in figure \ref{fig:VariableImportance} to predict $a_1$ for the unactuated case.
The distribution shows three regions of high relevance: $s\approx 0.25$, $s\approx 0.35$ (flap tip), and $s\approx 0.1$ in order of their importance. 
\section{\label{Sec:Num}Numerical setup}

The present investigations are based on three two-dimensional URANS numerical simulations of an airfoil equipped with a Coanda flap.
The test cases are selected for their different aerodynamic characteristics, which range from fully separated to nearly fully attached.
In the following, the airfoil configuration is introduced (\S~\ref{Sec:Config}), the numerical setup is detailed (\S~\ref{Sec:URANS}), 
and the different test cases are presented (\S~\ref{Sec:testcases}).

\subsection{\label{Sec:Config} Configuration}
\begin{figure}
\includegraphics[width=1\linewidth]{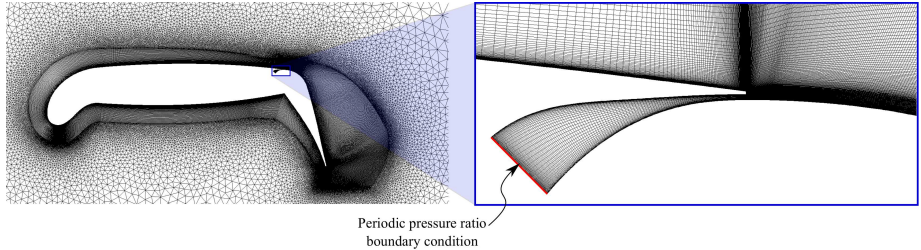}
\caption{\label{fig:configuration} Part of the numerical mesh surrounding the modified DLR F15 airfoil with a
close-up of the actuation duct.}
\end{figure}
The high-lift configuration, shown in figure \ref{fig:configuration}, is a modified DLR F15 airfoil equipped with a highly deflected Coanda flap and a droop nose. 
The details of the airfoil design are described in Burnazzi \& Radespiel  \cite{Burnazzi2014} and Jensch et al. \cite{jensch_07}, 
where the objective was to accomplish high lift coefficients during take-off and landing.
The leading edge geometry was reached after an iterative process that improved the airfoil stall behavior, which is ruled by the suction peak. 
The highly deflected flap at $65^{\circ}$ has a chord length of $c_{\rm{fl}}=0.25\,c$.
The numerical simulations are performed at Mach number $\mbox{Ma}=0.15$, and Reynolds number $\mbox{Re} = U_{\infty}c/\nu=12\cdot 10^6$, where $\nu$ is the kinematic viscosity of the fluid. 
These flow parameters correspond to the expected conditions during landing.

\subsection{\label{Sec:URANS}Unsteady Reynolds averaged Navier-Stokes simulations}

The computational fluid dynamics (CFD) solver employed to perform the analysis is the DLR TAU-Code  \citep{Kroll_02,Schwamborn_06}. 
The two-dimensional Unsteady Reynolds Averaged Navier-Stokes (URANS) equations are solved using a finite volume approach. 
The discretization schemes are the central scheme and the second order upwind Roe scheme for the mean-flow inviscid flux and the convective flux of the turbulence transport equation, respectively. 
The turbulence model is that of Spalart--Allmaras with curvature correction  \citep{Shur_00}, 
which allows the one-equation turbulence model to maintain a good accuracy in regions where the streamlines have high curvature. 
This characteristic is fundamental for the simulation of the Coanda phenomenon, 
which is based on the equilibrium between the inertial forces and the momentum transport in the direction normal to the convex surface  \citep{Pfingsten_07}. 
The numerical scheme and the turbulence model were previously assessed by comparing the results to wind tunnel experiments  \citep{Pfingsten_09,Pfingsten_09_2}. 
The lift, drag, and pitching moment coefficients are determined by integrating the pressure and shear stress distributions over the airfoil surface. 
The contribution from the added jet momentum is not included.
The periodic Coanda blowing mentioned in section \ref{sec:Intro} is performed by pulsating the pressure boundary condition at the base of the actuation chamber (see close-up in figure \ref{fig:configuration}).


The mesh is composed of a structured and an unstructured region, as the close-up in figure \ref{fig:configuration} shows.
The outer unstructured mesh has a C-block topology and extends 50 chord lengths in all directions.
The structured grid extends from the airfoil surface outward to cover the region where the main viscous phenomena occur. 
The viscous sub-layer is also resolved with $y_+<1$ everywhere over the airfoil surface. 
An important characteristic of the grid is the high density along the pressure side, where the stagnation point can be located, far from the leading edge. 
The structured region also extends over a large area behind the highly-deflected flap, in order to accurately capture the wake dynamics. 
Both, the trailing edge and the edge of the blowing slit, are discretized by means of a local C-block topology.

\subsection{\label{Sec:testcases}Test cases}
\begin{table}
\centering
\caption{\label{tab:testcases} Summary of the actuation conditions for the three test cases.}
\begin{tabular}{llll}
$F^+$ & $C_{\mu 0}$ & $C_{\mu 1}$ & Reference name\\
\hline
0 & 0 & 0 & Unactuated\\
1 & 0.011 & 0.0015 & Weakly actuated\\
1 & 0.035 & 0.0144 & Moderately actuated
\end{tabular}
\end{table}
In order to test the proposed method under different flow conditions, three test cases characterizing different flow separation states are numerically simulated. 
These test cases, summarized in table \ref{tab:testcases}, are characterized by three parameters, $C_{\mu 0}$, $C_{\mu 1}$ and $F^+$, 
of which only variations in $C_{\mu 0}$ and $C_{\mu 1}$ are considered. 
The non-dimensional actuation frequency $F^+$, defined as \citep{seifert1996,seifert:04a}
\begin{equation}
 F^+= \frac{f^a\>c_{\rm fl}}{U_{\infty}}\,,
\end{equation}
where $f^a$ is the actuation frequency, and $c_{\rm fl}$ is the flap chord length, is kept constant.
For both the weak and moderate blowing cases, it is set at $F^+=1$ (i.e. $f^a=204$~Hz), 
as recommended by Nishri and Wygnanski  \cite{Nishri1998}.
The total periodic blowing intensity (including both steady and unsteady components) can be prescribed as
\begin{equation}
 c_\mu(t)=\frac{{U}_{\rm J}(t)\>\dot{m_{\rm J}}(t)}{\frac{1}{2}\>\rho\> U_\infty^2\>S_{\rm{ref}}}\,,
\end{equation}
where $\dot{m}_{\rm J}$ is the jet mass flow rate, ${U}_{\rm J}$ is the jet averaged velocity across the slit, and $S_{\rm{ref}}$ is the reference area.
For the current periodic actuation, the momentum coefficient can be expressed as $c_\mu(t)=C_{\mu 0}+C_{\mu 1}\>\cos(2\>\pi\>f^a\>t)$,
where $C_{\mu 0}$ is the steady mean and $C_{\mu 1}$ is the oscillation amplitude.
This yields a range of mean momentum coefficients between $C_{\mu 0}=0$ and 0.035.
The oscillation amplitudes for the weakly and for the moderately actuated cases are $C_{\mu 1}=0.0015$ and 0.0144, respectively.
The angle of attack is kept constant at $0^{\circ}$ for all cases.

\section{\label{Sec:Results}Results and discussion}
The previous sections discussed the method and the numerical setup. 
In this section, the results for the circulation control wing are presented.
The powerful predictions of the machine learned models are first shown.
Optimal sensor positions are then presented for the different test cases, for the different state estimators and for two types of sensors.

\subsection{Unsteady flow estimation}
\begin{figure}
\includegraphics[width=\linewidth]{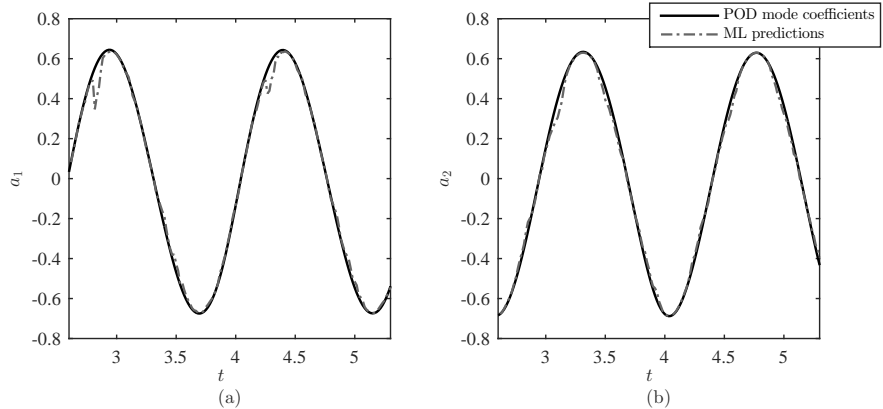}
\caption{\label{fig:LSEvsML} Reference POD (black), and machine-learning (gray) estimated mode coefficients using 
two optimal pressure sensors to estimate $a_1$ and $a_2$ for the unactuated case.}
\end{figure}

Before presenting the optimal sensor placement results, it is relevant to assess the prediction accuracy of the machine learned models upon which the method is based.
Figure \ref{fig:LSEvsML} presents the first two reference POD (black) and the predicted machine-learned (gray) mode coefficients using 
two optimal pressure sensors for the unactuated case.
The accuracy of the machine learned model is clear; the predictions follow the reference POD mode coefficients closely.
The small deviations and the overall prediction accuracy can be further improved by invoking additional techniques, 
such as sliding window methods or hidden Markov models \cite{Dietterich2002}.
Such techniques are not utilized in the current study, as the focus is on optimal sensor placement.

\begin{figure}
\centering
\begin{subfigure}{.5\textwidth}
  \centering
  \includegraphics[width=.9\linewidth]{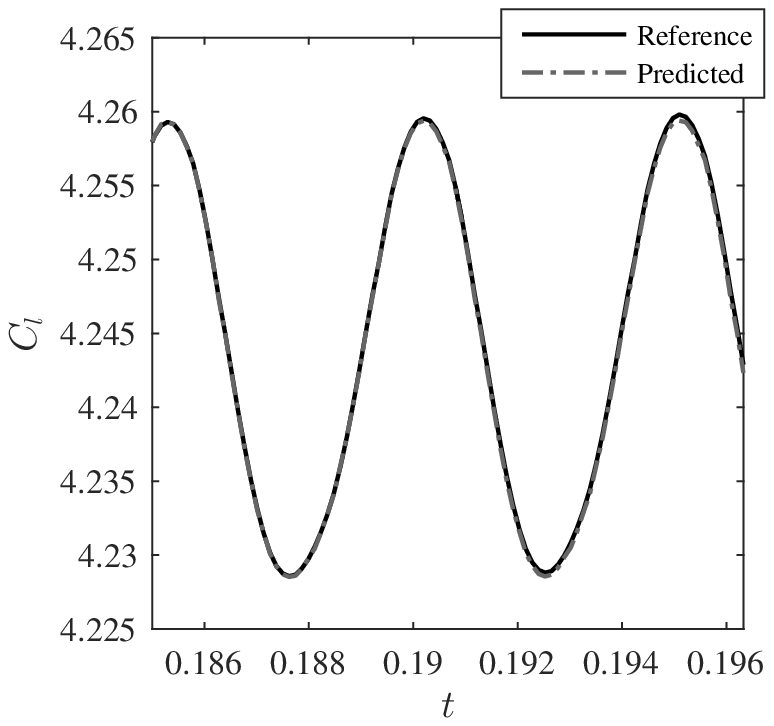}
  \caption{}
  \label{fig:Cl_Prediction}
\end{subfigure}%
\begin{subfigure}{.5\textwidth}
  \centering
  \includegraphics[width=.9\linewidth]{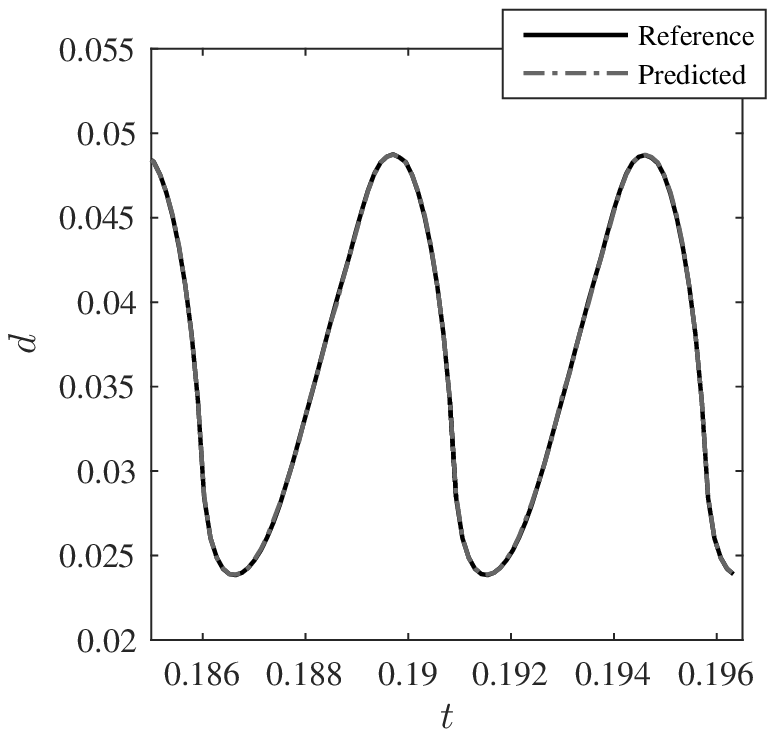}
  \caption{}
  \label{fig:d_Prediction}
\end{subfigure}
\caption{Reference (black) and estimated (gray) (a) lift coefficient $C_l$ and (b) streamline distance $d$ using two pressure sensors for the moderately actuated cases.}
\label{fig:ML_Predictions}
\end{figure}

Further validation of the models is presented in figure \ref{fig:ML_Predictions} for two different state estimators.
The figure shows the reference (black) and estimated (gray) (a) lift coefficient $C_l$ and (b) streamline distance $d$ using two optimal pressure sensors for the moderately actuated case.
The results are also in a very good agreement with the reference data.
The predictions are indiscernible from the URANS results.
Good agreement was also documented for other test cases and response functions. 

%
%

\subsection{Optimal sensor placement}

\subsubsection{\label{Sec:MethodValidation}Method validation}
\begin{figure}
\centering
\includegraphics[width=\linewidth]{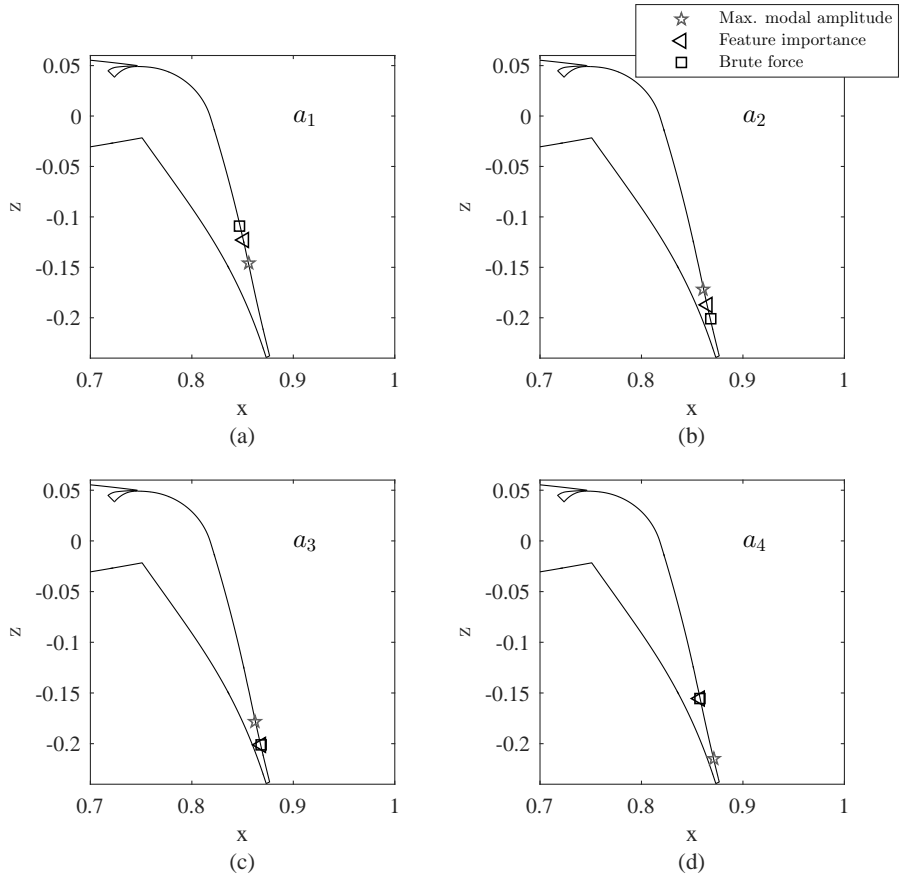}
\caption{\label{fig:ML_vs_POD} Optimal pressure sensor position to predict the first four POD mode coefficients for the unactuated case
 using the maximum modal surface value ($\pentagram$), the sensor with the highest feature importance ($\vartriangleleft$), and the brute force permutation ($\square$).}
\end{figure}

Whether heuristic \cite{Cohen2004} or mathematical \cite{Willcox2006, Yildirim2009},
POD-based approaches for optimal sensor placement positions the sensors at the extrema of the POD modes.
Since we have restricted the possible sensor positions to the aft surface of the airfoil, 
this method yields optimal locations at the maximum modal surface values \cite{Willcox2006}.
The results using this approach ($\pentagram$) are presented in figure \ref{fig:ML_vs_POD} alongside those using the proposed variable importance method ($\vartriangleleft$) for a single pressure sensor.
Also shown, are the optimal sensor positions using brute force permutations over all sensors ($\square$).
The brute force approach generates a machine learned model for each possible sensor (or for each possible sensor combination for multiple sensors configuration) and computes its OOB prediction error.
The optimal sensor location is then determined as the position with the lowest OOB error.
As the figure shows, the sensor positions using the variable importance method are always similar to those from the brute force approach.
This is not surprising since the bootstrapping in random forest is similar to the permutations in the brute force method.
Except for $a_4$, all three methods yield comparable sensor positions, thus providing confidence in the proposed method.
\subsubsection{Effect of response function choice and of flow conditions on optimal sensor placement}
\begin{figure}
\centering
\includegraphics[width= \linewidth]{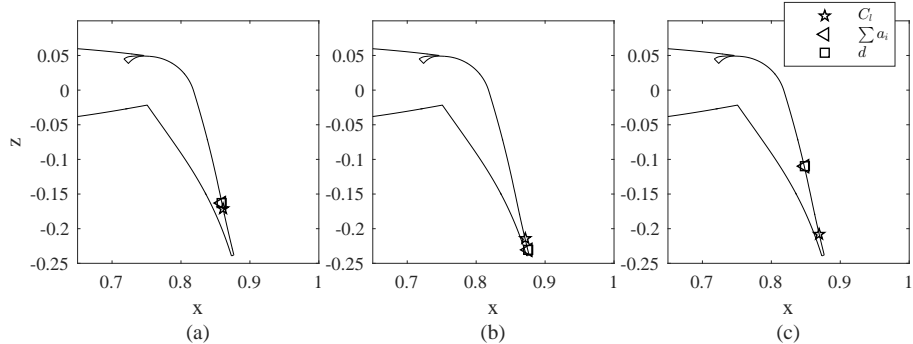}
\caption{\label{fig:OSP_AllStates} Optimal pressure sensor position for the three response functions for the (a) unactuated, 
(b) the weakly actuated and (c) the moderately actuated test cases.}
\end{figure}

As introduced in section \ref{sec:ResponseFunction}, the state of the flow can be inferred using various metrics.
The effect of three different response functions on the optimal pressure sensor positions for the three test cases is illustrated in figure \ref{fig:OSP_AllStates}.
The symbol $\sum a_i$ designates the sum of the first 4 POD mode coefficients, whereas $C_l$ and $d$ denote the lift coefficient and the streamline distance, respectively.
As the figure shows, the sensor positions using $\sum a_i$ and $d$ are surprisingly analogous for all three cases.
This suggests that the streamline distance delivers similar information about the instantaneous flow state as the POD mode coefficients.
The sensor locations using $C_l$ differ slightly than those from the other two response functions, especially for the moderately actuated case.

\begin{figure}
\centering
\includegraphics[width = 6cm]{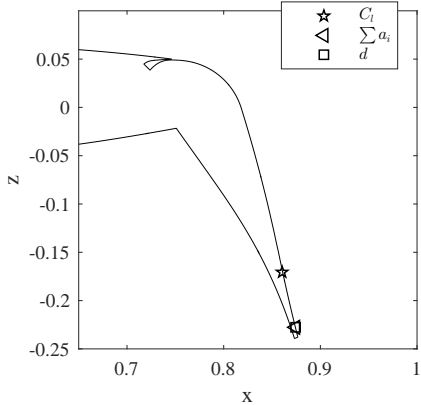}
\caption{\label{fig:OSP_Combined} Optimal pressure sensor position for the three response functions for the combined test case.}
\end{figure}

The effect flow conditions have on the optimal sensor position can be also observed in figure \ref{fig:OSP_AllStates}.
As the flow gradually changes from fully separated to almost fully-attached, 
the optimal sensor position also changes.
For all cases, the sensor appears to favors locations closer to the trailing edge than to the flap shoulder.
A clear trend in the position evolution is however difficult to discern. 
Such change in optimal sensor positions is clearly problematic for closed-loop flow control experiments,
which experience a range of conditions.
A compromise that best captures the flow state under the various conditions is therefore necessary.
This compromise can be obtained by combining the data from the three test cases (using the same record length for each case) into a single dataset and repeating the same procedure as before. 
Alternatively, one can perform one long simulation with varying forcing conditions (such as with chirp forcing).
Figure \ref{fig:OSP_Combined} presents the optimal pressure sensor position for the three response functions for the combined test case.
The sensor positions using $\sum a_i$ and $d$ are again similar.
They are located near the tip, which is similar to the weakly actuated case.
The sensor location using $C_l$ again differs from the results of the two other response functions.
It is located at approximately the same location as for the unactuated case.

\subsubsection{Effect of sensor type on optimal sensor placement}
\begin{figure}
\centering
\includegraphics[width=\linewidth]{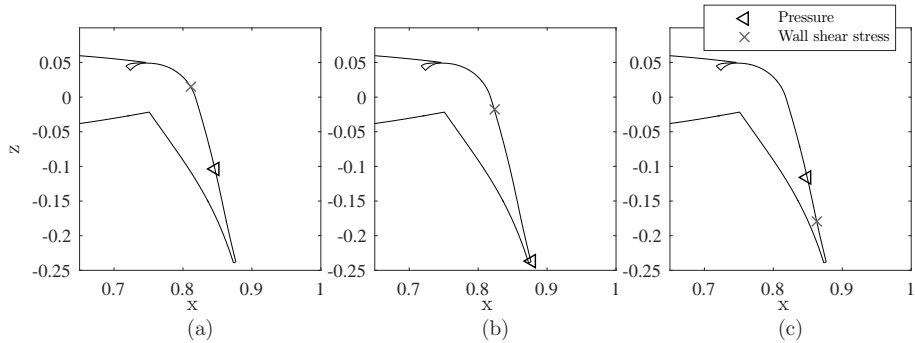}
\caption{\label{fig:OptSens_P_Cf} Optimal sensor position to estimate $a_2$ using two sensor types for the (a) unactuated, (b) the weakly actuated, (c) and the moderately actuated test cases.}
\end{figure}

The sensor type plays an important role in an experiment, whether to understand the flow behavior or to use as input for closed-loop control.
Typical sensors for aerodynamic applications measure either pressure or shear stress.
In order to simulate the signal of an experimental hot-film sensor which is insensitive to the flow direction, 
the absolute value of the numerical wall shear stress data was initially used.
This step, however, was superfluous as the results between the sensor locations using the raw and the absolute values were similar.
Figure \ref{fig:OptSens_P_Cf} presents the optimal sensor positions to estimate $a_2$ using either a pressure or a wall shear stress sensor 
for the (a) unactuated, (b) the weakly actuated, (c) and the moderately actuated test cases.
As the figure shows, the optimal sensor position clearly depends on the sensor type.
For all three test cases, the pressure sensor location shifts between the middle and tip of the flap.
On the other hand, the optimal shear stress sensor is located around the flap shoulder for the unactuated case and gradually shifts towards the tip with higher actuation intensity.
This difference between the two sensor types can be attributed to the different flow phenomena they respectively capture;
the wall shear stress sensor appears to be best positioned at the start of the separation region, 
whereas the optimal pressure sensor is located at the maximal modal value over the flap surface.
The optimal wall shear stress sensor location does not seem to be correlated with the POD extrema.
This obviously has important consequences on the design of POD model-based closed-loop control experiments using hot-films.


\section{\label{Sec:Conclusions}Conclusions}

In this study a new machine learning-based method for optimal sensor placement is introduced.
The method first constructs machine learned models from a range of input sensors to predict a response function.
The optimal sensor positions are then determined as the most relevant features (input) through variable importance ranking. 
The method offers several advantages over existing alternatives: it is simple, adaptable, and computationally inexpensive,
e.g. one POD computation (excluding additional processes such as Gappy POD) of 200 snapshots requires 309 seconds on a single Intel Core i7-4770 CPU with a $3.40\>$GHz processor, 
whereas training the machine learning model and determining the variable ranking takes only 17.2 seconds for the same record.
The machine learning algorithm used for this study is random forest, due to its accuracy and inherent randomization.
However, any supervised machine leading regression algorithm is suitable.

The proposed approach is implemented on a circulation control wing for a range of actuations conditions.
The analysis is based on three two-dimensional unsteady Reynolds averaged Navier-Stokes (URANS) simulations of the airfoil, which is equipped with a drooped nose and with a Coanda flap.
The three simulations are conducted at different actuation frequencies and blowing intensities, yielding different flow conditions.

The model prediction accuracy is validated through the low out-of-bag error and by comparing the model output to reference data.
Good agreement between the model predictions and the reference results is observed for three response functions: the POD mode amplitudes, the lift coefficient, and the streamline distance.
Using the variable importance ranking, the optimal sensor positions are determined.
The results are compared against the extrema modal amplitude and against a brute force permutations approach for the first four POD mode amplitudes.
The three methods yield comparable sensor positions, providing confidence in the proposed approach.

The effect of the response function choice on the sensor placement is examined.
The results suggest that as long as the response function is physical and reflect the state of the flow, 
its type has little influence on the optimal sensor positions.
On the other hand, flow conditions clearly affect the sensor placement.
For a flow with varying conditions, we recommend training data that simulate the entire range of expected conditions.
The optimal sensor position would then be a compromise over that entire range.
The type of sensor has also an effect on the sensor positioning.
Recommendations concerning the type of sensor are strongly related to the experiment's objectives.
If the objective is solely the physical understanding of the flow, then either pressure or hot-film sensors can be used.
If the objective is closed-loop flow control, then a pressure sensor might be better suited since its sensitivity to varying flow conditions appears to be lower.
The recommendation for pressure sensor stems also from a practical perspective, as hot-film sensors are usually difficult to calibrate.

To the best of our knowledge, the present study is the first to introduce a machine-learning based approach for optimal sensor placement. 
The proposed method can be employed in any fluid dynamics application and even in other fields such as structural health monitoring.
It can be exploited to design better experiments and control laws.

\section*{Acknowledgements}
We acknowledge the funding and excellent working conditions 
of the Collaborative Research Centre (CRC 880) 
`Fundamentals of High Lift of Future Civil Aircraft'
supported by the Deutsche Forschungsgemeinschaft (DFG) 
and hosted at the Technical University of Braunschweig.

\section*{References}

\bibliography{CF_Ref}

\end{document}